# Carrier-Coupled Ultrafast Structural Dynamics and Interlayer Energy Transport of Supported Transition Metal Dichalcogenide Heterostructures

Md Shaikot Alam Shakil[1], Ting-Hsuan Wu[1], Xing He[1], Abu Montakim Tareq[1], Zhenjia Zhou[2], Libo Gao[2], Naihao Chiang[1], Ding-Shyue Yang[1,*]

[1]*Department of Chemistry, University of Houston, Houston, Texas 77204, United States*

[2]*National Laboratory of Solid State Microstructures, Jiangsu Key Laboratory for Nanotechnology, School of Physics, Collaborative Innovation Center of Advanced Microstructures, Nanjing University, Nanjing 210093, China*

*To whom correspondence should be addressed. Email: yang@uh.edu

**ABSTRACT**

Understanding the electronic coupling and energy flow across layered two-dimensional heterostructures (HSs) is crucial to the exploitation of carrier and phonon transports as well as thermal management in next-generation optoelectronic devices. By using reflection ultrafast electron diffraction, we directly examine photoinduced out-of-plane structural dynamics of supported $MoS_2/WS_2$ bilayer HSs and their individual monolayers. Experimental evidence reveals the launch of ultrafast carrier-coupled intralayer atomic motions due to interlayer charge transfer across the van der Waals (vdW) heterojunctions that is absent for individual monolayers. Such a notable carrier–lattice correlation is in addition to the electronic coupling manifested in the enhanced optical absorption for HSs. Also, different pathways of energy flow as a result of carrier–phonon coupling and phonon scattering are reported with the corresponding characteristic times. On longer timescales, relaxation of thermalized atomic motions can be sufficiently described by a thermal transport model. A higher thermal boundary conductance (TBC) across $MoS_2/WS_2$ HSs is obtained compared to those at the monolayer–substrate interfaces; however, the similar TBC values suggest comparable couplings of phonons across vdW contacts. These results further shed light on the optical, phonon, and interfacial thermal properties of vertically-stacked vdW HSs.

Van der Waals (vdW) heterostructures (HSs) formed by vertically stacking atomically thin two-dimensional (2D) materials provide a versatile platform to explore and engineer emergent electronic, optical, and thermal properties.[1,2] The pronounced anisotropy, exemplified by strong in-plane (IP) covalent bonds and comparatively weak out-of-plane (OP) interlayer interactions, enables chemical or mechanical strategies to form diverse materials combinations for atomically defined interfaces, with reduced sensitivity to lattice-matching constraints compared to conventional systems.[1–3] Such versatility has enabled the realization of artificial functional materials with desired band alignments and correlated and/or interfacial phenomena. Factors such as dielectric screening, proximity effects, and interlayer hybridization, can further influence the electronic structure in a manner sensitive to the stacking configuration and thickness and to external perturbations.[4–6] These features establish vdW HSs as model systems in which electronic structures and lattice degrees of freedom are intrinsically intertwined across atomically thin interfaces.

Dynamically, the electronic landscape of 2D HSs can be shaped with ultrafast photoexcitation. For example, in type-II HSs composed of semiconducting transition metal dichalcogenides (TMDCs), interlayer charge transfer (ICT) takes place within 100 femtoseconds (fs), with rapid redistribution of photogenerated carriers across adjacent layers.[7–9] Depending on the excitation level, these photocarriers may either form Coulomb-bound interlayer excitons with spatial charge separation and prolonged lifetimes,[10–13] or evolve into an electron–hole plasma when the density exceeds the Mott threshold.[14,15] In the latter regime, enhanced Coulomb screening leads to band renormalization, photoluminescence quenching, and spectral broadening. Consequently, the ultrafast redistribution of charged carriers also modifies the interlayer energy landscape, thereby perturbing the interactions and forces acting on the lattice ions across the vdW gap. Previous experimental investigations employing Raman,[16,17] transient absorption[18,19] and reflectivity,[20] and terahertz emission spectroscopy[21] have revealed that phonon-mediated energy relaxation and interlayer interactions are governed by carrier–phonon coupling and play key roles in the nonequilibrium dynamics of vdW HSs. Therefore, transient electronic perturbations can also induce structural responses as a result of the electronic–lattice correlation. On longer timescales, it is technologically important to understand how excess energy transferred to the lattice dissipates across the HS and into the supporting substrate.[22–24] Compared to strong IP chemical bonds and therefore high thermal conductivities, weak interlayer vdW interactions and the mismatch of vibrational density of states may be limiting factors for the coupling of OP phonon modes, hence suppressing energy transfer and giving rise to substantial thermal boundary resistance in artificially stacked vdW assemblies.[23,25] Thus, it is crucial to have atomic-level spatiotemporal resolution to directly probe various photoinitiated structural responses.

In recent years, time-resolved diffraction methods, especially ultrafast electron diffraction (UED), have been employed to study nonequilibrium electronic–lattice responses and structural dynamics of layered vdW systems such as TMDCs,[26–33] graphene,[34,35] black phosphorus,[36,37] topological insulators,[38,39] SnSe,[40] MXenes,[41] etc. Many of these reports were conducted in transmission geometry, which directly probes IP photodynamical phenomena that may be used to infer OP information when multilayers or multiple components are involved. For TMDC HS,[31,42–44] photoinduced ICT has been shown to generate rapid phonon emission and nearly simultaneous lattice heating in both constituent layers, revealing strongly coupled carrier–phonon interactions that redistribute energy across atomically smooth interfaces. Recently, coherent torsional lattice motions driven by carrier redistribution have been reported for twisted bilayers by examining electron scattering signals for collective restructuring dynamics of moiré superlattices.[44] However, direct probing of ultrafast OP atomic motions and dynamical interlayer coupling in HSs remains limited given the transmission geometry used.

In this work, we employ UED in reflection geometry to directly monitor the OP mean-square atomic displacements of epitaxially gown $MoS_2/WS_2$ HSs on sapphire(0001)[45] and their constituent monolayers following photoexcitation, which informs the electronic–lattice coupling, cross-plane lattice thermalization, and interfacial heat transport. A distinct structural response within the instrumental response time is observed due to ultrafast ICT, which is absent in the behavior of individual monolayers. Given the different characteristic times for TMDC monolayers, a physical picture emerges for the energy flow pathways among the subsystems of photocarriers, IP phonons, and OP atomic motions, where carrier-coupled ultrafast OP structural motions become an additional energy flow route for HSs. By quantitatively extracting the temporal evolution of lattice motions and modeling interfacial heat diffusion, we further determine the cross-plane thermal boundary conductances (TBCs) of supported $MoS_2/WS_2$ HSs and their constituent monolayers; the similar TBC values suggest comparable couplings of phonons across vdW contacts. These results demonstrate that vdW stacking does not merely superimpose individual layer dynamics but fundamentally reshapes ultrafast photodynamics including structural motions and interfacial energy flow through coupled electronic and phononic interactions.

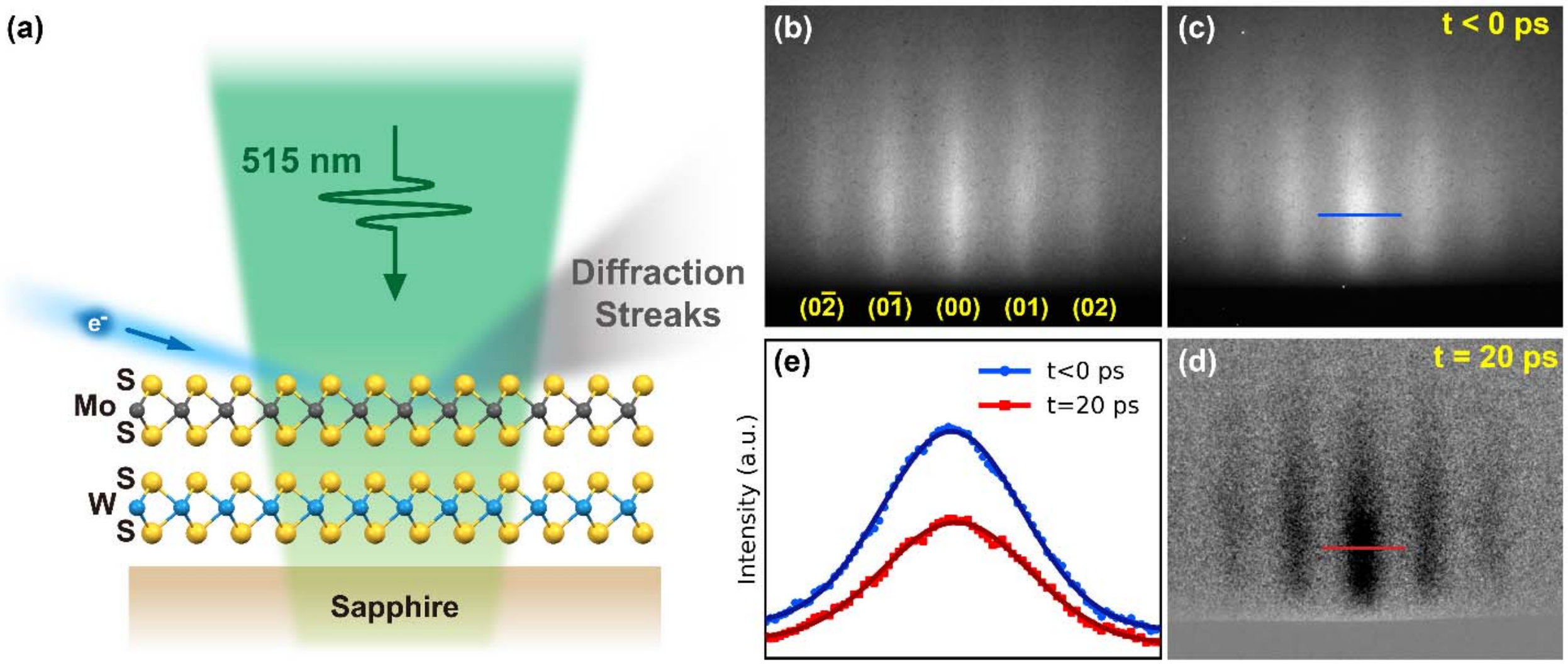

**Figure 1.** Schematic of reflection UED experiments and diffraction images. (a) A schematic showing the optical excitation of sapphire-supported $MoS_2/WS_2$ by a 515-nm pulsed laser and probed by 30-keV electrons at a grazing incidence of ~2.4°. (b) Ordered diffraction streaks with the Miller indices acquired from the top 2D $MoS_2$ layer of the HS using a parallel electron beam. (c) Broadened diffraction streaks as a result of the use of a gently focused electron beam. The image is a reference frame acquired prior to photoexcitation ($t < 0$). (d) Diffraction differences observed at 20 ps with a laser fluence of 3.2 mJ $cm^{-2}$. (e) Diffraction intensity profiles of the center streak along the horizontal blue and red lines indicated in panels c and d for $t < 0$ and $t =$ 20 ps, respectively. The solid curves are fits to a Gaussian function with a linear background.

## Results and Discussion

### Grazing-Incidence Electron Diffraction of Supported TMDC Samples

Shown in Figure 1a is a schematic of the reflection UED measurements conducted on sapphire-supported $MoS_2/WS_2$ HS bilayer and individual monolayer (1L) TMDC samples, where 515-nm light is used for photoexcitation and 30-keV electrons probe the OP structural dynamics at a grazing incidence of $\theta_{\text{in}} \cong 2.4°$. Given the electron elastic mean free path of $l \cong 108$ Å at 30 keV,[28] the probe beam is predominantly sensitive to the upper $MoS_2$ layer of the HS ($MoS_2^{HS}$) as the single-scattering depth of $l \cdot \sin\theta_{\text{in}} \sim 4.5$ Å is comparable to the thickness of 6–7 Å for a TMDC monolayer. Prior to photoexcitation, a single-zone pattern of ordered diffraction streaks is observed from $MoS_2^{HS}$ as well as separate monolayers (Figure 1, b and c, and Fig. S1), which confirms the long-range crystallinity of all the TMDC samples fabricated on sapphire(0001).[45] The high sample homogeneity is also evidenced by the uniform Raman spectra with the characteristic HS and 1L peaks observed at select locations across the entire sample areas (Fig. S2). Upon illumination by 515-nm light, both $MoS_2^{HS}$ and $WS_2^{HS}$ in the HS are photoexcited and fast intensity decreases are observed for all diffraction streaks of $MoS_2^{HS}$ (Figure 1d). The time-dependent changes are quantitatively extracted from fits of the horizontal intensity profiles of the center streak with a Gaussian function (Figure 1e). The same experimental conditions are also applied to sapphire-supported 1L $MoS_2$ and $WS_2$ samples in order to compare their photoinitiated dynamics.

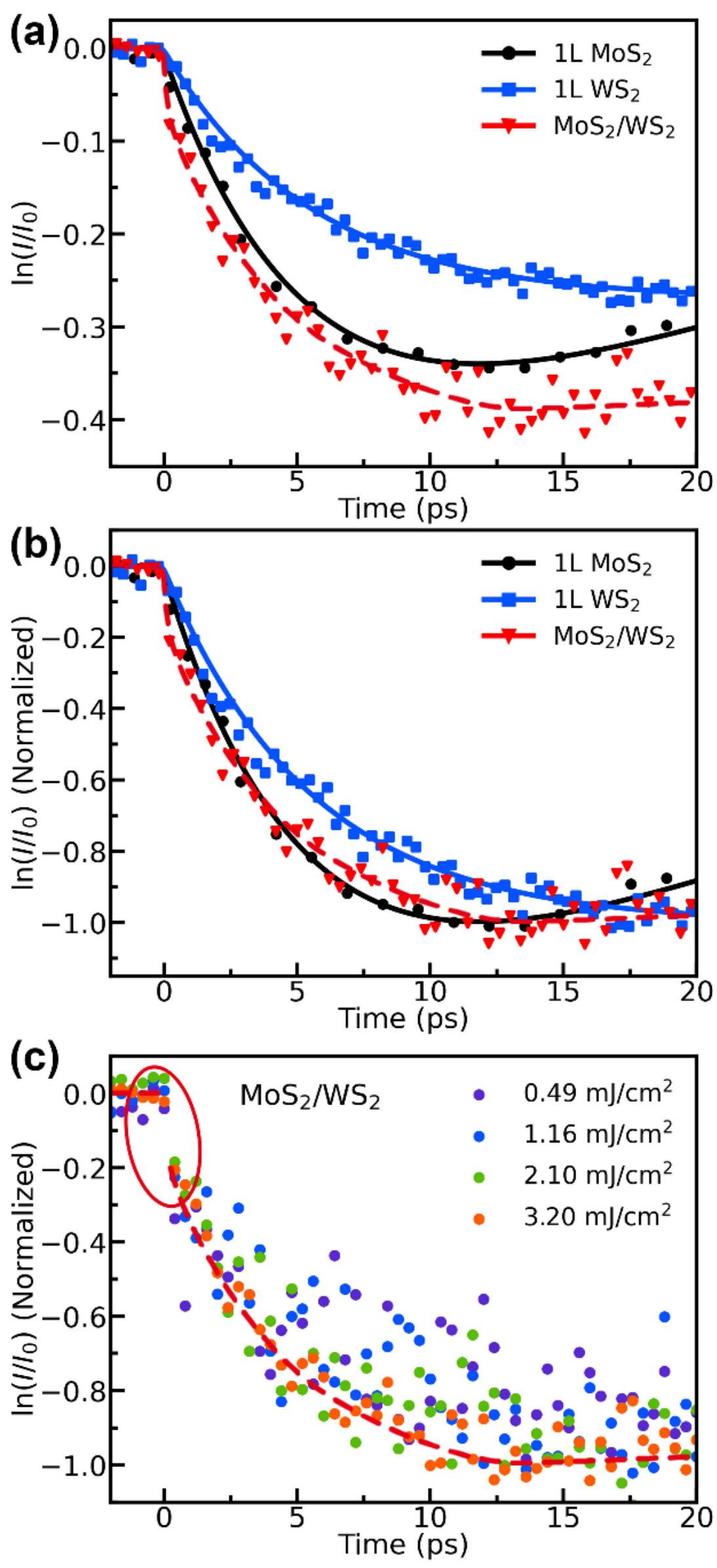

**Figure 2.** Photoinitiated OP structural dynamics at early times. (a) Time-resolved diffraction intensity changes of $MoS_2/WS_2$ HS at 3.2 mJ $cm^{-2}$ compared to those of 1L $MoS_2$ and $WS_2$. The black and blue solid lines are fits to a function with coupled exponential rise and recovery. The red dashed line is a guide to the eye. (b) Normalized time-dependent changes with observation respect to the corresponding maximum changes. (c) Normalized dynamics of $MoS_2^{HS}$ measured at select fluences. The red dashed line is a guide to the eye based on the data at 3.2 mJ $cm^{-2}$.

**Out-of-plane Structural Dynamics at Early Times**

Shown in Figure 2a is a comparison of the early-time diffraction intensity changes at 3.2 mJ $cm^{-2}$, which covers the temporal range from the initial ultrafast response to lattice thermalization before or at around 20 ps. Three major behavioral differences are notable. First, $MoS_2^{HS}$ as part of the HS exhibits an enhanced overall diffraction intensity decrease of ~47.5%, appreciably more than those of 1L $MoS_2$ (~41%) and $WS_2$ (~33%) when the layers are separate. This observation indicates a direct consequence of increased photoabsorption as a result of significant interlayer interactions, which are not present in individual 1Ls, across the vdW contact.[18,46] Second, the rise times to reach the respective maximum decreases for the OP direction are prominently different, especially between 1L $MoS_2$ and $WS_2$ (Fig. 2b); however, we note that their IP structural dynamics have been shown by transmission UED to be largely the same.[31,42] Interestingly, the diffraction intensity decrease of $MoS_2^{HS}$ is somewhat slowed down in the first 10-ps window compared to 1L $MoS_2$. Third, and most importantly, $MoS_2^{HS}$ exhibits an ultrafast photoinduced response of ~20% of the overall change within a single step of 400 fs, which is lacking in the 1L photodynamics (see also Fig. S3). Such a 20% ultrafast change within the instrumental response time is consistently seen for different laser fluences used in this study (Fig. 2c). We also note that the complex $MoS_2^{HS}$ photodynamics cannot be described by a single exponential rise function, whereas the early-time UED results of the individual monolayers can.

The fluence-dependent results show further HS-specific features of the photoinitiated dynamics in addition to the apparent linear responses (Fig. 3, a and b, based on the results shown in Fig. S4). For the maximum diffraction decrease reached, $MoS_2^{HS}$ exhibits a more negative linear slope than the individual monolayers, which is consistent with the observed larger decrease for the HS as a result of enhanced photoabsorption illustrated in Fig. 2a. Strikingly, a non-zero intercept is found for $MoS_2^{HS}$, which is absent for the monolayer samples (Fig. 3b). Such a nonzero intercept together with the aforementioned few-hundreds-of-fs ultrafast component points to the presence of an ultrafast structural response in the OP direction unique to the HS; this intercept is also reminiscent of the reported light-induced stress at low excitation densities in TMDCs where an apparent nonzero intercept can be seen by extrapolating the higher-fluence data.[47] It is known that charge transfer takes place in type-II HSs on an ultrashort time scale.[7,8,18] Thus, our observation provides direct evidence for ultrafast carrier-coupled structural dynamics across the stacked layers, which we anticipate to be more prominent because of the band-alignment-guided ICT compared to those of multilayer TMDCs.[47] The impact of charge transfer on the UED dynamics can also be seen in the rise time constants within the first 20 ps (Fig. 3c). While 1L $MoS_2$ exhibits a rise time of $\tau_{r,M}^{\perp} \sim 2.5$ ps at laser fluence of ≥0.5 mJ $cm^{-2}$ (for photoinjection of an electron–hole plasma above the Mott density), $MoS_2^{HS}$ in the HS displays a notably larger apparent rise time of $\tau_{r,HS}^{\perp} \sim 6$–7 ps; intriguingly, 1L $WS_2$ shows an even larger rise time of $\tau_{r,W}^{\perp} \sim 7.5$ ps for the OP dynamics. When the laser fluence is further reduced, all time constants increase but the HS shows the slowest rise among all TMDC samples studied (Fig. 3c).

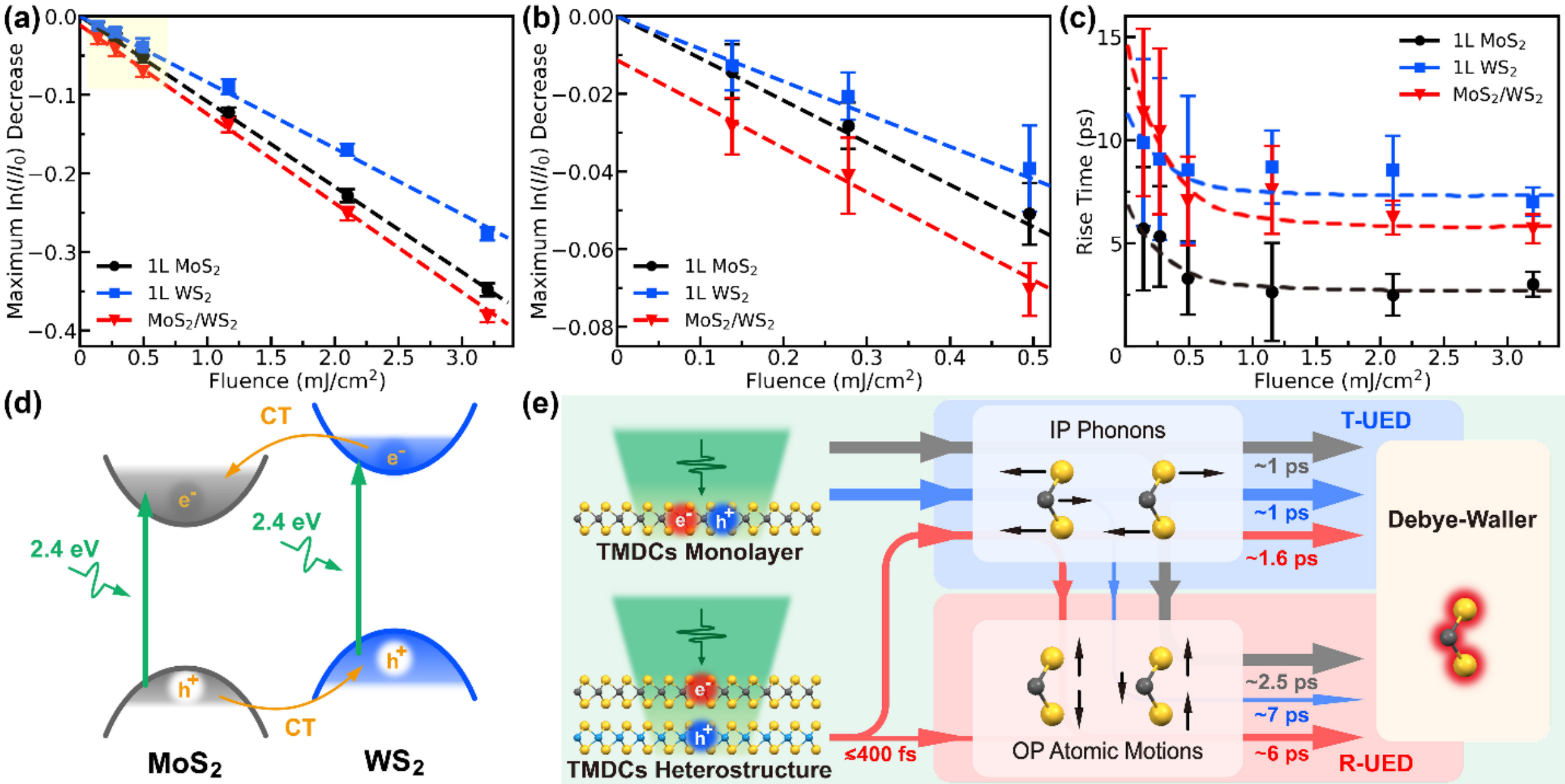

**Figure 3.** Fluence dependence of diffraction changes and photoinduced processes in TMDC materials. (a) Maximum diffraction intensity decreases as a function of photoexcitation fluence for 1L $MoS_2$, 1L $WS_2$, and $MoS_2/WS_2$ HSs. Dashed lines are linear fits of the data. (b) Expanded view of the low-fluence range from panel a. (c) Apparent rise time constants $\tau_r^{\perp}$ of the individual monolayers and the HS as a function of laser fluence. Dashed lines are guides to the eye. (d) Type-II band alignment and ultrafast ICT in $MoS_2/WS_2$ HSs. (e) Schematic of photoinjection of charged carriers (left), carrier–phonon coupling, major phonon modes involved (middle, with black arrows indicating the atomic motions), and phonon thermalization pathways in TMDC

monolayers (gray and blue lines/arrows for $MoS_2$ and $WS_2$, respectively) and electronically-coupled HSs (red lines/arrows). The associated time constants are denoted, together with different thicknesses of the lines/arrows representing the rates inversely corresponding with the times. The upper blue and lower red boxes indicate the most sensitive structural probing for UED in transmission (T) and reflection (R) geometries, respectively.

In reflection UED, unless there are directional atomic or optical phonon motions,[36,48] a steady reduction in diffraction intensity especially on a time scale of a few to tens of ps can be associated with an increase in the average OP mean-square atomic displacements $\Delta\langle u_\perp^2(t)\rangle$ according to the Debye–Waller relation,[28]

$$\ln\frac{I(t)}{I_0} = -4\pi^2 s_\perp^2\,\Delta\langle u_\perp^2(t)\rangle \tag{1}$$

where $I_0$ and $I(t)$ denote the diffraction intensities of the center streak before photoexcitation and at time $t$, respectively, and $s_\perp$ is the OP momentum transfer. Based on the maximum changes in Fig. 2a that represent thermalized atomic motions, we obtain $\Delta\langle u_\perp^2\rangle_{\max}$ to be about 0.010 $Å^2$, 0.0088 $Å^2$, and 0.0068 $Å^2$ for $MoS_2^{HS}$, 1L $MoS_2$, and 1L $WS_2$, respectively. (See Fig. S5 for the effects of the *intralayer* $A_1'$ and $A_2''$ phonons, the two modes with vertical atomic motions, on reflection diffraction intensity. We rule out the scenario of dominant $A_1'$ or $A_2''$ phonon motions at early times.) The increased OP motions are the results of carrier–phonon coupling, carrier annihilation, and phonon–phonon scattering to reach lattice thermalization, on time scales determined by TMDC materials' dynamic properties. For monolayers, the lifetime of a photoinjected electron–hole plasma has been shown to be as short as <1 ps.[28,49–52] Thus, given photocarriers' stronger coupling to the IP phonons, this signifies a comparable time constant $\tau_r^{\parallel}$ for the rise of IP atomic motions due to energy transfer and phonon scattering, which has been reported for TMDC 1Ls and HSs by transmission UED and time-resolved grazing-incidence x-ray diffraction studies.[27,42,44,53–55] The value of $\tau_r^{\parallel}$ is almost the same for 1L $MoS_2$ (~1.7 ps)[53,55] and $WS_2$ (~1.6 ps).[42] The observation of $\tau_{r,M}^{\perp}$ ~ 2.5 ps being close to $\tau_r^{\parallel}$ indicates that phonon scattering and thermalization in all directions are fast for 1L $MoS_2$, which is consistent with its near-isotropic nature of atomic motions reported earlier.[28] In contrast, the prominent difference between $\tau_r^{\parallel}$ and $\tau_{r,W}^{\perp}$ for 1L $WS_2$ informs a highly anisotropic lattice response and thermalization bottleneck, which is reminiscent of the behavior of a similar TMDC.[54]

However, a markedly different temporal evolution of the HS UED results beyond a single exponential-rise response points to unique photodynamics. The electronic origin is illustrated in Fig. 3d, where appreciable photoinjected electrons (holes) are transferred to $MoS_2^{HS}$ ($WS_2^{HS}$) and a spatial interlayer charge separation is reached on an ultrafast time. Thus, the HS introduces an additional nonequilibrium pathway driven by ICT (Fig. 3e, lower) compared to TMDC monolayers, where the structural responses are primarily governed by the coupling between carriers and IP phonons and the following phonon thermalization (Fig. 3e, upper). Our observation of an ultrafast OP structural response of $MoS_2^{HS}$ with a residual zero-fluence intercept is consistent with the phonon-assisted ICT picture given in earlier theoretical reports, showing particularly the major role of the $A_1'$ optical phonon mode.[56,57] On a qualitative level, the spatial separation of electrons and holes establishes a transient Coulombic attraction perpendicular to the interface and hence an interlayer force on the atoms. Such OP interlayer interactions can also couple to the lattice degrees of freedom for coherent breathing motions[20] or for moiré superlattice reconstruction in twisted HSs.[44] However, due to the fast phonon–phonon scattering, long-lasting intralayer coherent $A_1'$ oscillations are not detected by our method (as shown in Fig. S5a, the structure factor simulations give no diffraction decrease upon average over an $A_1'$ oscillation cycle). Instead, the effective outcome is an ultrafast incoherent energy transfer pathway opened up for the HS, in addition to the relatively longer-time IP couplings, from the electronic subsystem to the OP atomic motions. We also note that the compressive strain is not directly resolved because of the limited probe depth of 30-keV electrons and therefore the observation of diffraction streaks without clear spacing information.

Thus, the schematic in Fig. 3e summarizes the energy flow pathways experimentally resolved by UED in two geometries. For monolayers, relaxation of the photoinjected electron–hole plasma in the same layer takes place on a sub-ps timescale (~1 ps) by coupling to IP phonons predominantly, which results in an lattice response governed by IP phonon scattering and thermalization (Fig. 3e, horizontal arrows in the upper panel). Further scattering and thermalization with the OP degrees of freedom (vertical arrows between the middle upper and lower panels in Fig. 3e) leads to a rise of OP atomic motions. The coupling of IP and OP motions determines the intrinsic time for energy redistribution, with $\tau_{\mathrm{r,M}}^{\perp}$ ~ 2.5 ps for 1L $MoS_2$ and $\tau_{\mathrm{r,W}}^{\perp}$ ~ 7.5 ps for 1L $WS_2$ observed in reflection UED; the different thicknesses of the lines in Fig. 3e signify the different rates. In contrast, the HS exhibits a branched relaxation pathway that involves both intralayer and interlayer channels under the effects of ICT (Fig. 3e, lower panel).

Given the photoinjection level, a large portion of the photoexcited carriers still relax via intralayer IP mechanisms resembling the monolayer behavior with a similar time. However, phonon-assisted ultrafast ICT activates a pathway unique to the HS that leads to OP atomic motions and hence an early-time non-thermal structural response. In addition, an increase in the carrier lifetime due to ICT prolongs the recombination and consequently the energy transfer to phonons. Thus, compared to $\tau_{\mathrm{r,M}}^{\perp} \sim 2.5$ ps for 1L $MoS_2$, a larger time constant of $\tau_{\mathrm{r,HS}}^{\perp} \sim 6$ ps from $MoS_2^{HS}$ is observed. In about 12–25 ps, the lattice subsystem ultimately reaches thermalization, and afterward the diffraction intensity changes are well described by temperature jumps under the Debye–Waller framework (Fig. 3e, left).

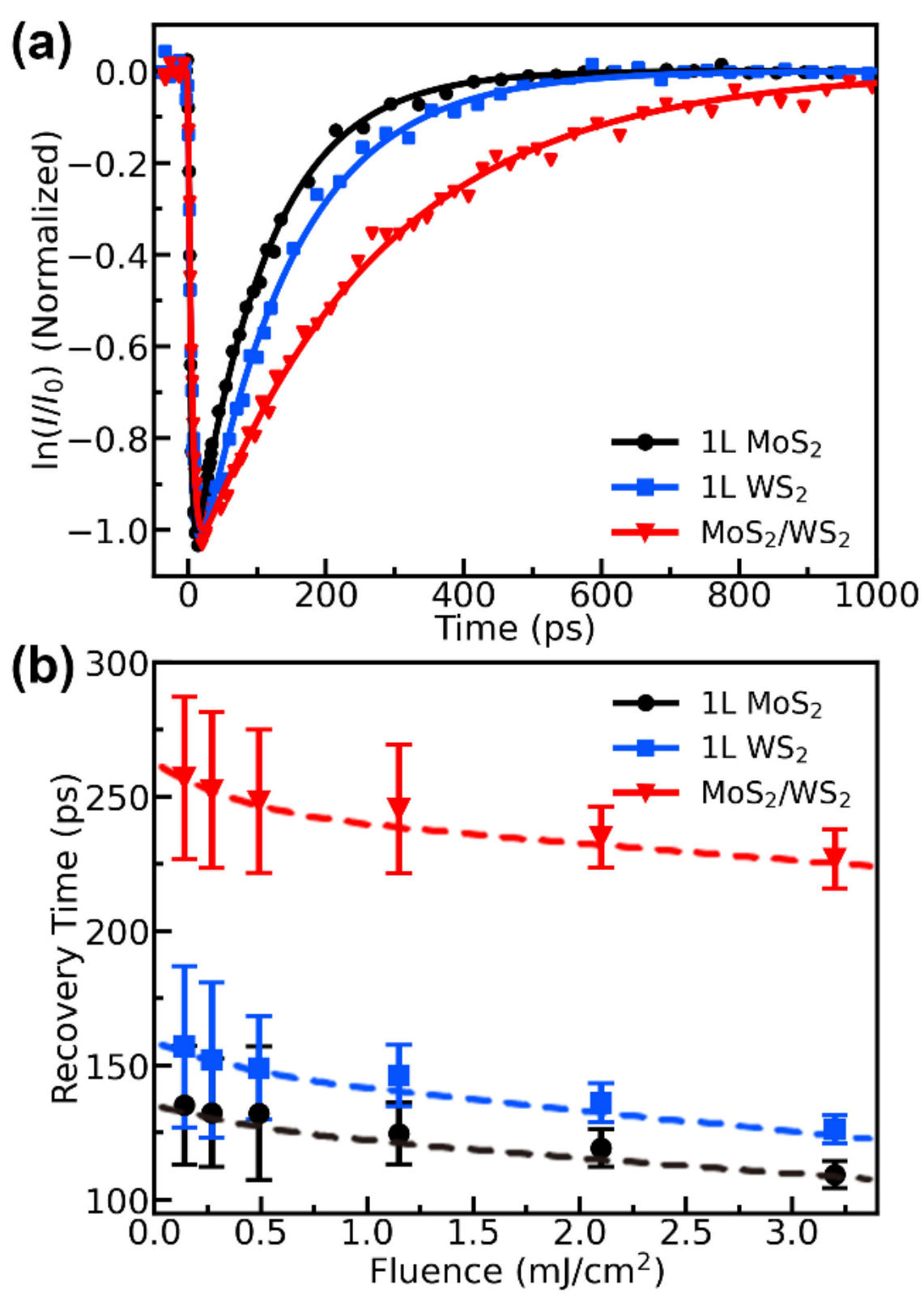

**Figure 4.** Structural recovery dynamics of TMDC monolayers and HSs at longer times. (a) Time-dependent recovery of diffraction intensities of 1L $MoS_2$, 1L $WS_2$, and the $MoS_2/WS_2$ HS measured at 3.2 mJ $cm^{-2}$. (b) Apparent recovery time constants for the individual monolayers and the HS as a function of excitation fluence. Dashed lines are guides to the eye.

### Interlayer and Interfacial Thermal Transport at Longer Times

We now turn our attention to the recovery of laser-heated TMDCs via thermal dissipation toward the supporting substrate. While 1L $MoS_2$ and $WS_2$ exhibit comparable thermal decay times of $\tau_{\mathrm{d,M}}$ = 110 ± 5 ps and $\tau_{\mathrm{d,W}}$ = 130 ± 7 ps, respectively, $MoS_2^{HS}$ in the HS shows notably slowed recovery with a time constant of $\tau_{\mathrm{d,HS}}$ = 220 ± 12 ps at 3.2 mJ cm$^{-2}$ (Fig. 4a). These apparent recovery times slightly increase at lower excitation fluences but the overall trend remains. The result of significantly slower recovery for $MoS_2^{HS}$ than 1L $MoS_2$ agrees with a similar finding from a photoexcited $WSe_2/WS_2$ heterojunction by transmission UED.[42] We also note that $\tau_{\mathrm{d,HS}}$ falls nicely in the interlayer thermalization time range of 190 and 350 ps obtained by molecular dynamics assuming local intralayer thermal equilibrium.[31]

To quantitatively rationalize the slow recovery of HSs, a one-dimensional thermal transport model is constructed by considering the sapphire-supported $MoS_2/WS_2$ HS (1L $MoS_2$ and $WS_2$) as a 3-component (2-component) thermally-coupled system, with an effective temperature representing each component and a TBC for each interface. The experimentally obtained $\Delta\langle u_\perp^2(t)\rangle$ (Eq. 1) can be converted into a corresponding lattice temperature $T(t)$ according to the Debye model,[58]

$$4\pi^2\Delta\langle u_\perp^2(t)\rangle \approx \frac{3h^2}{\bar{m}k_B\Theta_D^2}\left[T(t)-T_0\right] \quad (2)$$

where $h$ is the Planck constant, $\bar{m}$ the average atomic mass, $k_{\mathrm{B}}$ the Boltzmann constant, $\Theta_D$ = 580 K (460 K) the Debye temperature of 1L $MoS_2$ ($WS_2$),[28,59] and $T_0$ = 295 K is the base temperature (Fig. 5, a to c). Thus, the maximum temperature jump $\Delta T_{\max}$ at each fluence $F$ is derived as the initial condition. Details about the model can be found in Supplementary Information. It is worth noting that $\Delta T_{\max}$ and $F$ follow a linear relation considering complete thermalization: $\Delta T_{\max} = F\chi/C\rho L$, where $C$, $\rho$, and $L$ ~ 0.7 nm are the specific heat, mass density, and thickness of a TMDC, respectively, and $\chi$ is the absorption coefficient. Hence, from the linear slopes, we obtain the $\chi$ values of 3.9%, 2.9%, and 4.4% at 515 nm for supported 1L $MoS_2$, 1L $WS_2$, and $MoS_2^{HS}$ in the HS, respectively (Fig. 5a–c, insets). It is further noted that $MoS_2^{HS}$ in the HS exhibiting larger photoabsorption than standalone 1L $MoS_2$ cannot be attributed to the result of classical Fresnel equations for a multilayer summation of reflections and absorption (see Fig. S6 and Supplementary Information for further discussion). We emphasize the notable impacts of electronic coupling on the photodynamics of HSs beyond the behavior of individual layers.

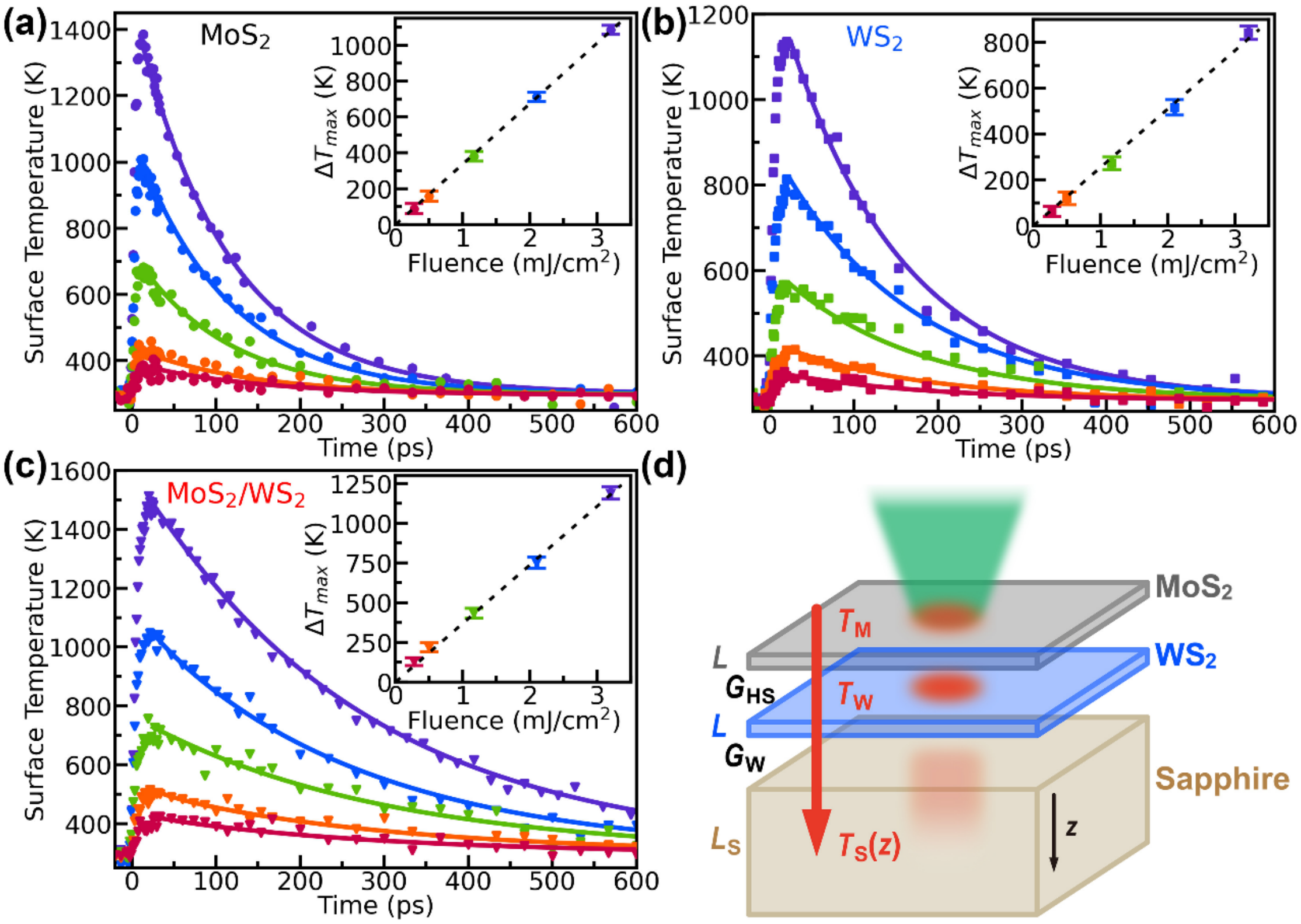

**Figure 5.** Comparison of experimentally-derived temperature jumps with the thermal transport model of TMDC systems. (a–c) Effective lattice temperatures derived from the UED data (symbols) over excitation fluences ranging from 0.27 to 3.2 mJ cm$^{-2}$ for (a) 1L $MoS_2$, (b) 1L $WS_2$, and (c) $MoS_2^{HS}$ in the HS. Solid lines are theoretical results. Insets display the corresponding maximum temperature jumps at select fluences, with the standard deviations indicated, as well as the linear dependence (dashed line). (d) Schematic of cross-plane heat dissipation in the supported HS, illustrating the layer thicknesses, the corresponding temperatures, and the two interfacial TBCs.

For 1L $MoS_2$ and $WS_2$ on sapphire, we reproduce the experimental recoveries at all excitation fluences with the TBCs of $G_M \sim 13$ MW m$^{-2}$ K$^{-1}$ and $G_W \sim 10.5$ MW m$^{-2}$ K$^{-1}$, respectively (Fig. 5, a and b), which signify a weaker interfacial heat transfer for $WS_2$ compared to $MoS_2$. These values can be quantitatively validated by the relation $\tau_d \cong C\rho L/G$, because the little temperature jumps at the supporting sapphire surface (Figs. 5d and S7b) effectively reduce the coupled thermal transport to a first-order heat-loss differential equation.[28] We then extend the thermal model to the $MoS_2/WS_2$ HSs, with the $WS_2$–sapphire TBC being fixed at the determined $G_W$. An interlayer TBC of $G_{HS} \sim 25$ MW m$^{-2}$ K$^{-1}$ is obtained at the $MoS_2$–$WS_2$ heterojunction, which indicates a comparably more efficient heat transfer across the HS than across either monolayer–substrate interface. Considering the multiple components, the total cross-plane effective TBC $G_{\mathrm{eff}}$ can be expressed as[60]

$$\frac{1}{G_{\mathrm{eff}}} = \frac{L}{\kappa_{\mathrm{M}}} + \frac{1}{G_{\mathrm{HS}}} + \frac{L}{\kappa_{\mathrm{W}}} + \frac{1}{G_{\mathrm{W}}} + \frac{L_{\mathrm{S}}}{\kappa_{\mathrm{S}}}, \quad (3)$$

where $L$ is the thickness, $\kappa$ is thermal conductivity, and the subscripts M, W, HS, and S stand for $MoS_2$, $WS_2$, HS, and sapphire, respectively. Consequently, we obtain an effective $G_{\mathrm{eff}}$ = 6.6 MW $m^{-2}$ $K^{-1}$ for the multilayer stack, which yields an excellent agreement of $\tau_d$ = 220 ps for the recovery of $MoS_2^{HS}$ on the top.

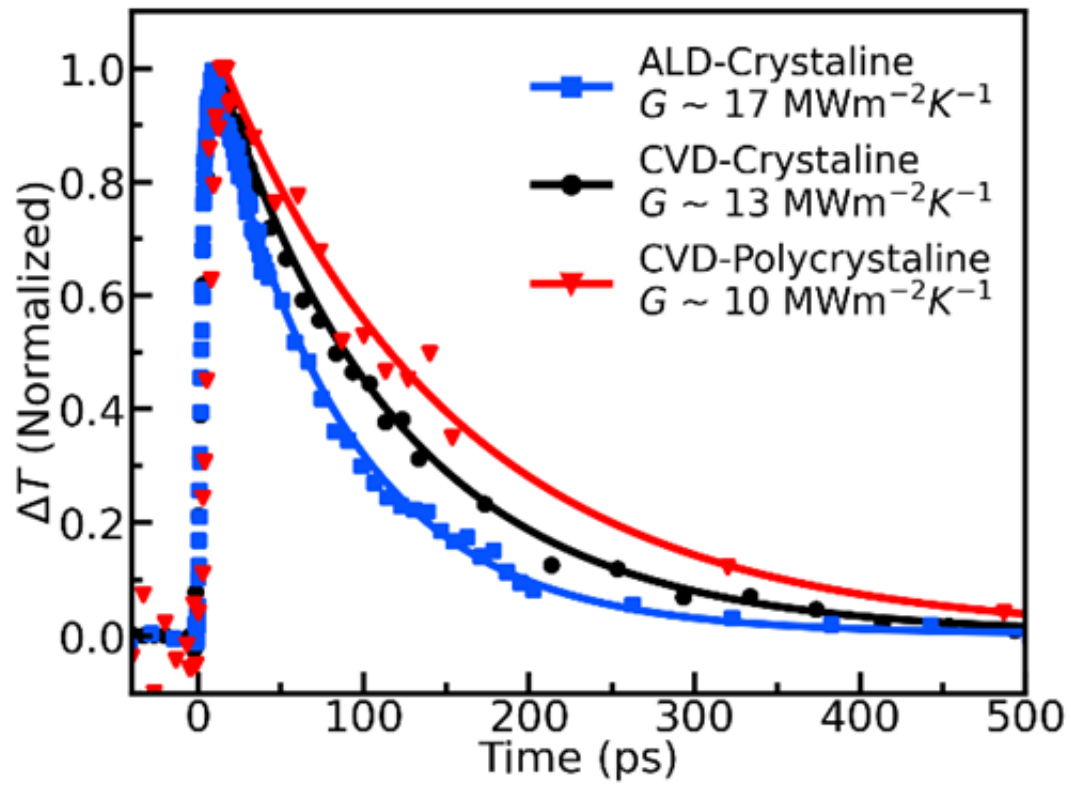

**Figure 6.** Comparison of the recovery dynamics of photoexcited $MoS_2$ fabricated as atomic layer deposition (ALD)–grown crystalline (blue; this work), chemical vapor deposition (CVD)–grown crystalline (black; this work), and CVD-grown polycrystalline (red; adapted from[28]) films. Solid lines represent fits using the thermal transport model.

Lastly, we compare the TBC values experimentally found for different TMDC vdW interfaces. It has been reported that deposited metal–nonmetal interfaces exhibit TBCs in the range of 50 to 500 MW $m^{-2}$ $K^{-1}$, and the interaction strength across a vdW gap between like materials can also have major effects.[61] The 2D/2D TBC has been reported to be lower than the 2D interface TBC with three-dimensional substrates by using Raman thermometry for equilibrium-style measurements.[25] However, our results show that the $MoS_2/WS_2$ interface TBC is actually higher than others as well as those previously reported (9 MW $m^{-2}$ $K^{-1}$ reported between $MoS_2$ and $WSe_2$[25] or 5 MW $m^{-2}$ $K^{-1}$ between $WSe_2$ and $WS_2$[42]), which signifies better phonon coupling that may be attributed to the use of the same chalcogen and a good lattice match. For the $MoS_2$/sapphire interface, we further find that the TBC varies between ~10 to ~17 MW $m^{-2}$ $K^{-1}$ depending on the fabrication method and sample crystallinity (Figure 6). These samples were prepared by large-area chemical synthesis with technological relevance instead of small-area transfer. The fact that the TBC values reported here are all within a relatively narrow range indicates the level of cross-plane and interfacial heat dissipation involving vdW-layered TMDC materials. However, the higher TBC obtained between $MoS_2$ and sapphire signifies better phonon coupling at the interface and hence more efficient acoustic phonon transmission likely as a result of a careful treatment of the substrate surface prior to growth of a high-quality crystalline film and post-growth annealing. The lowest TBC obtained from a polycrystalline film may be attributed to the growth condition, grain-boundary scattering, and less ideal interfacial contact. Thus, the results here illustrate the importance of carefully examining engineering methods to control the conditions of interlayer stacking and vdW contacts even for thermal management purposes in 2D optoelectronic devices.

**Conclusions**

In summary, we have employed UED in reflection geometry to directly probe photoinitiated OP structural dynamics of $MoS_2/WS_2$ bilayers and individual monolayers and reveal the impacts of interlayer charge transfer and energy transport across HSs. Pronounced ultrafast intralayer OP atomic motions in the HS are observed within the instrumental response time as a result of their strong coupling to ultrafast interlayer charge redistribution. Furthermore, the electronic coupling between HS layers across the vdW gap leads to higher photoabsorption beyond the sum of the constituent monolayers and modifies the thermalization of OP motions. Hence, we obtain the overall energy flow pathways and their characteristic times for the TMDCs studied. At longer times, the relaxation dynamics are described well by multi-component interfacial thermal transport. A higher thermal boundary conductance is found between $MoS_2/WS_2$ bilayers than those between individual monolayers and the supporting sapphire substrate. Fabrication of TMDCs and sample crystallinity are also found to have notable effects on interfacial thermal resistivity. However, the still comparable cross-plane heat flow signifies a potential bottleneck in interfacial energy transport in TMDC vdW systems. Taken together, the results of this study describe photoinduced structural dynamics of layered HSs over an optical cycle, providing a framework that links electronic excitation, atomic motions, and interfacial heat flow and offering insights on carrier–phonon coupling and thermal transport for considering the incorporation of TMDCs in next-generation optoelectronic and energy devices.

## Methods

Crystalline $MoS_2$ and $WS_2$ monolayers and $MoS_2/WS_2$ bilayer HSs were fabricated on sapphire (0001) by chemical vapor deposition.[45] The high crystallinity and uniformity of the large-area samples were confirmed by electron diffraction and Raman spectroscopy. Samples prepared by atomic layer deposition (ALD) followed the procedures in Ref.[62]

Details of reflection UED measurements have been described previously.[28,36] In short, photoexcitation was made by 515 nm light generated via second harmonic generation (SHG) of the fundamental output (1030 nm) of a Yb:KGW regenerative amplifier with a pulse duration of 170 fs and a repetition rate of 10 kHz. The optical pump beam passed through a beam-front tilt setup to minimize the velocity mismatch issue for reflection probing. A second stage of SHG of the 515-nm light produced 257-nm ultraviolet pulses for photoelectron generation from a $LaB_6$ emitter at 30 keV. The electron probe pulses were directed onto the sample surface at a grazing incidence angle of ~3.8°. All measurements were conducted at a base temperature of 295 K. Diffraction images were collected by a phosphor screen coupled to a gated intensified CMOS camera system. The instrumental response time of the reflection UED apparatus can reach a few-100s-of-fs level with the use of a reduced number of <100 electrons per pulse, which enables the observation of early-time ultrafast structural dynamics. To improve the signal-to-noise ratio and maintain a reasonable acquisition time, several hundred electrons per pulse were typically used for measurements on ps-to-ns timescales, which results in an instrumental response time of approximately 2 ps.

ASSOCIATED CONTENT

**Supporting Information**

Characterizations of TMDC samples, ultrafast OP structural responses of the $MoS_2/WS_2$ HS within the instrumental response time, fluence-dependent UED dynamics, dependence of diffraction intensities on OP $A_1'$ or $A_2''$ motions, analysis of optical absorption based on classical Fresnel equations for a multilayer summation of reflections and absorption, thermal transport model for relaxation of laser-heated TMDC systems.

AUTHOR INFORMATION

**Corresponding Author**

*Email: yang@uh.edu    Phone: +1 713-743-6022

**ORCID**

Ding-Shyue Yang: 0000-0003-2713-9128

Naihao Chiang: 0000-0003-3782-6546

Md Shaikot Alam Shakil: 0009-0002-8807-871X

Ting-Hsuan Wu: 0000-0001-5055-304X

Xing He: 0000-0001-5341-5662

Abu Montakim Tareq: 0000-0003-2704-7610

**Author Contributions**

The manuscript was written through the contributions of all authors. All authors have given approval to the final version of the manuscript.

**Notes**

The authors declare no competing financial interest.

**Acknowledgments**

We thank Drs. Devendra Pareek and Sascha Schäfer for the ALD samples. S.A.S., T.-H.W., X.H., and D.-S.Y. acknowledge the main support by the R. A. Welch Foundation (E-2263) and partial support for the instruments from the National Science Foundation (CHE-2154363). A.M.T. and N.C. acknowledge support from the National Science Foundation (CHE-2304955).

# Supporting Information

# Carrier-Coupled Ultrafast Structural Dynamics and Interlayer Energy Transport of Supported Transition Metal Dichalcogenide Heterostructures

Md Shaikot Alam Shakil[1], Ting-Hsuan Wu[1], Xing He[1], Abu Montakim Tareq[1], Zhenjia Zhou[2], Libo Gao[2], Naihao Chiang[1], Ding-Shyue Yang[1,*]

[1]*Department of Chemistry, University of Houston, Houston, Texas 77204, United States*

[2]*National Laboratory of Solid State Microstructures, Jiangsu Key Laboratory for Nanotechnology, School of Physics, Collaborative Innovation Center of Advanced Microstructures, Nanjing University, Nanjing 210093, China*

*To whom correspondence should be addressed. Email: yang@uh.edu

## S1. Characterizations of sapphire-supported TMDC samples

Shown in Fig. S1 are the electron diffraction images of epitaxially grown monolayer (1L) $MoS_2$ and $WS_2$ on sapphire(0001) recorded at two different zones, which shows their high crystallinity and azimuthal orientation order with respect to the substrate surface. Measurements of Raman spectroscopy were performed at room temperature with a confocal Raman microscope (WITec alpha300) equipped using a 1200 grooves/mm grating and a dry 100× (NA=0.9) objective to further confirm the uniformity of the large-area TMDCs monolayers and heterostructure (HS). A 532-nm laser with a power of approximately 1 mW was used. Shown in Fig. S2a are the comparison of the characteristic peaks of 1L $MoS_2$, 1L $WS_2$, and the $MoS_2/WS_2$ HS. The in-plane (IP) $E'$ and out-of-plane (OP) $A'$ modes at 385 and 404 $cm^{-1}$, respectively, with a peak separation of ~19 $cm^{-1}$ confirm the monolayer nature of 1L $MoS_2$.[S1] The $E'$ and $A'$ modes at around 353 and 416 $cm^{-1}$, respectively, are consistent with 1L $WS_2$.[S2] Hence, coexistence of the characteristic Raman peaks from both $MoS_2$ and $WS_2$ confirms the vertical stacking of the bilayer HS, where the observed slight shifts and spectral broadening relative to the isolated monolayers are the results of interlayer coupling.[S3] Closely resembling Raman spectra collected from five randomly selected positions confirm the spatial homogeneity of all samples used in this study (Fig. S2, b to d).

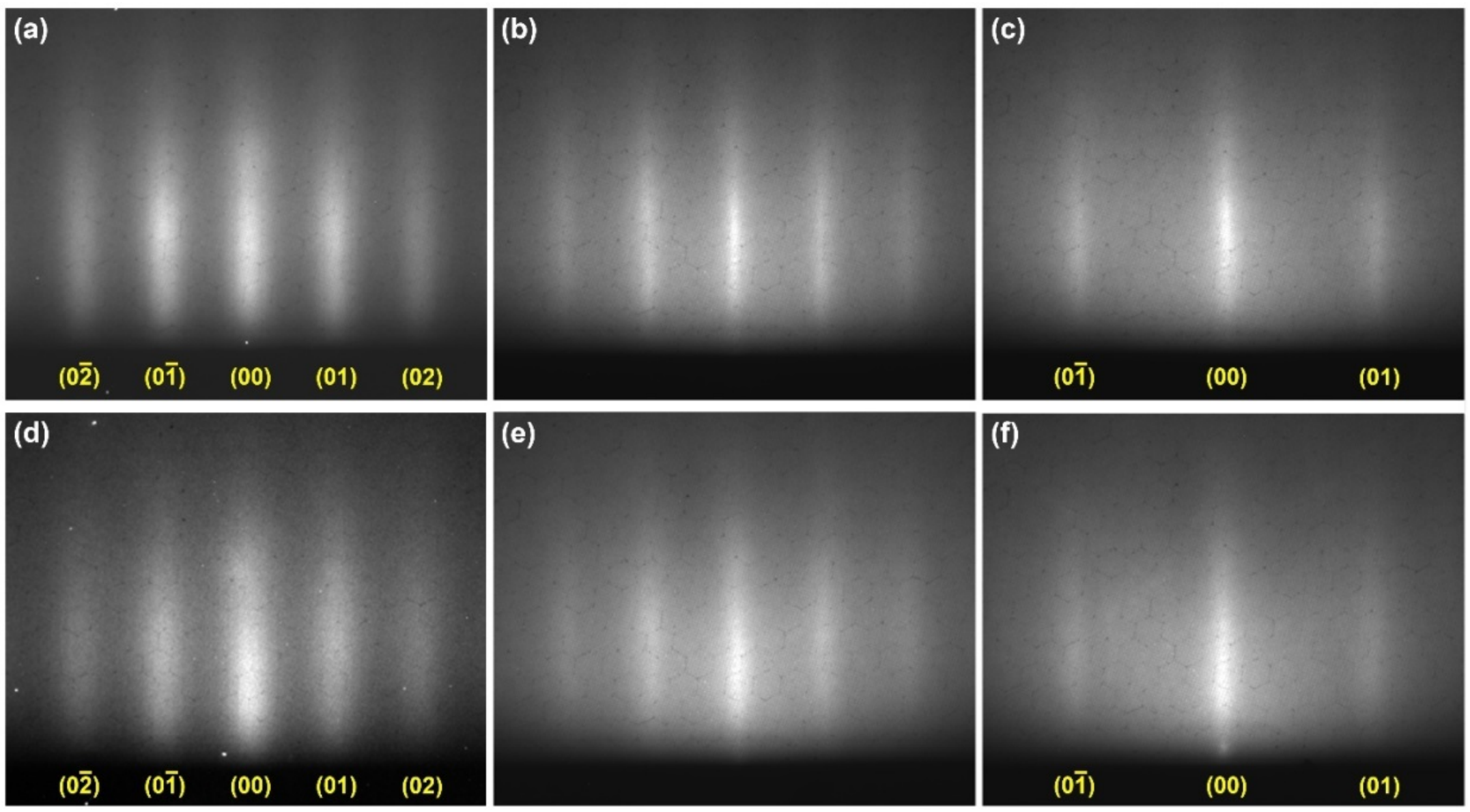

**Figure S1**: Reflection electron diffraction images of (a–c) 1L $MoS_2$ and (d–f) 1L $WS_2$, with the Miller indices specified for the observed streaks. A gently focused electron beam was used for panels a and d, where a parallel beam used for other panels yields sharper diffraction streaks with narrower horizontal widths.

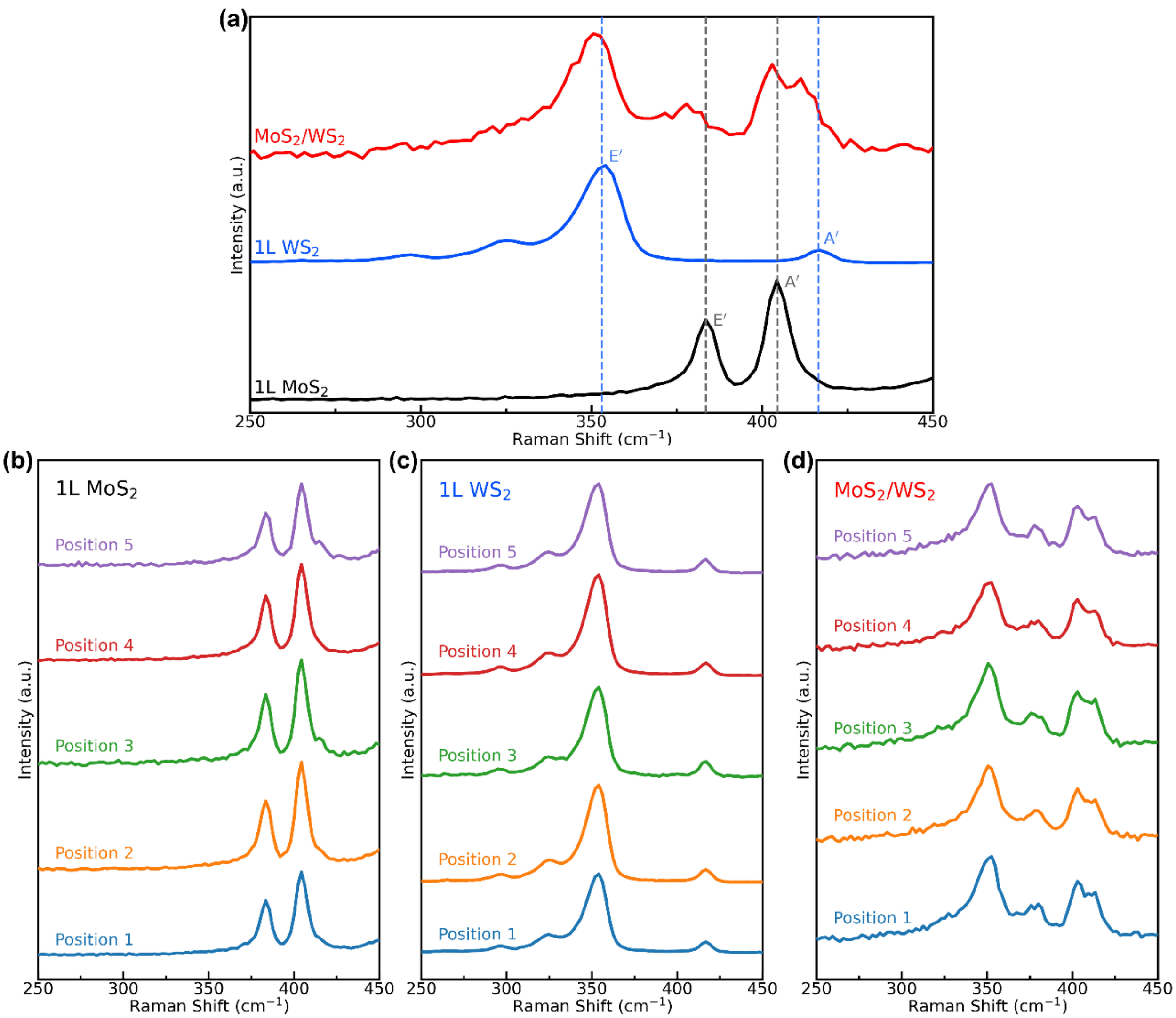

**Figure S2**: Raman spectra of the TMDC monolayers and bilayer HS obtained at room temperature. (a) Comparison of the spectra for 1L $MoS_2$ (black), $WS_2$ (blue), and the $MoS_2/WS_2$ HS (red). Highly resembling Raman spectra were obtained from five positions randomly selected across the large areas of (b) 1L $MoS_2$, (c) 1L $WS_2$, and (d) the $MoS_2/WS_2$ HS.

## S2. Ultrafast OP structural responses of the $MoS_2/WS_2$ HS

Shown in Fig. S3 are the dynamics of the $MoS_2/WS_2$ HS at early times recorded using two different step sizes (200 and 400 fs) for the delays between photoexcitation and electron probing. The abrupt OP structural response immediately following photoexcitation is ultrafast within the instrumental response time of a few hundred fs.[S4] Further diffraction intensity decreases in a more gradual, steady manner as a result of carrier–phonon coupling and phonon scattering/thermalization.

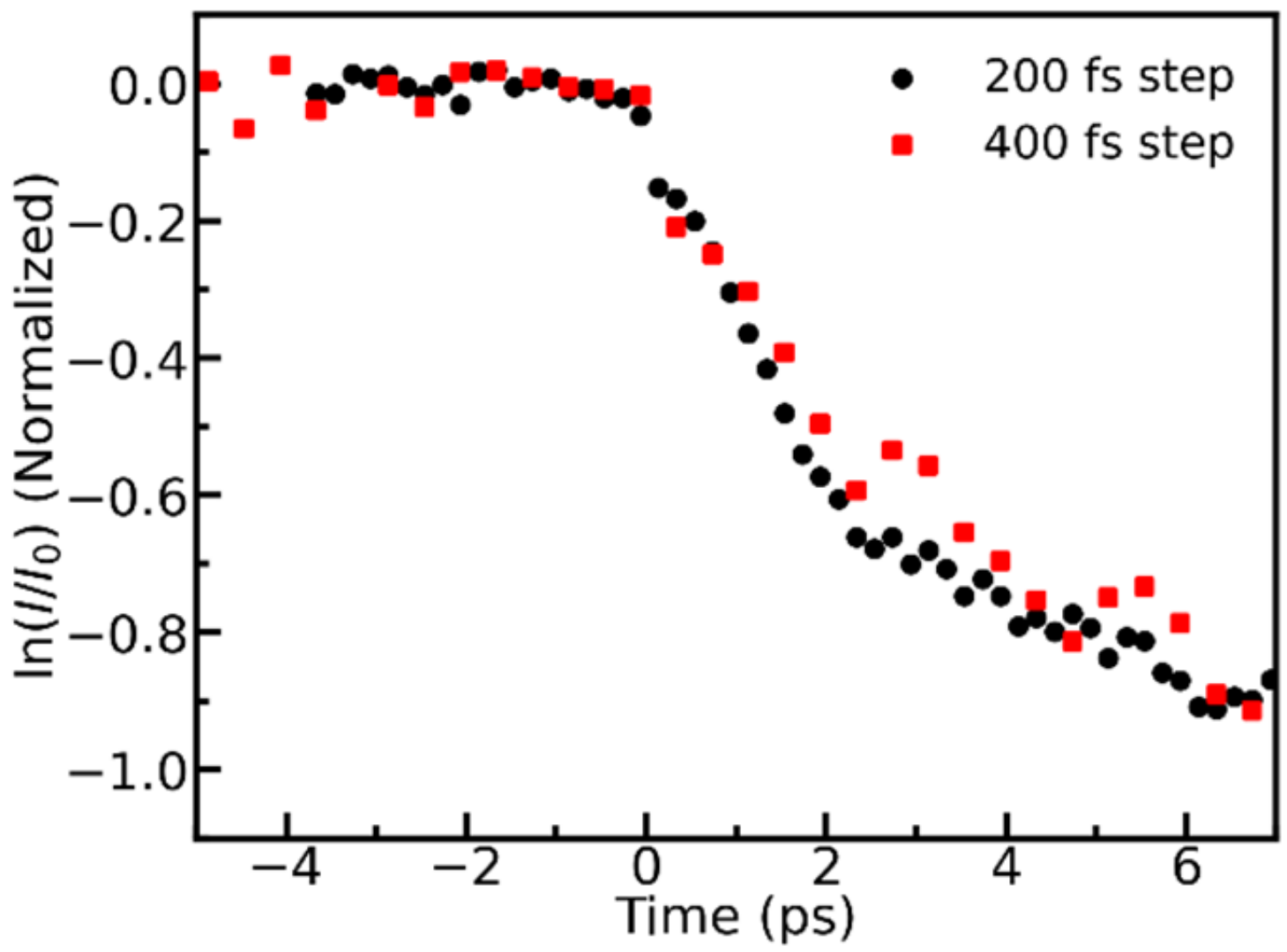

**Figure S3**: Ultrafast photoinduced responses of $MoS_2/WS_2$ HS recorded by reflection UED at early times, where ~20% of the overall change is seen initially by using the step sizes of 200 fs (black) and 400 fs (red) for pump–probe delays.

## S3. Fluence-dependent UED dynamics of TMDC monolayers and HSs

Shown in Fig. S4 are the overall UED dynamics of the TMDC monolayers and HS recorded at select fluences ranging from 0.14 to 3.2 mJ cm$^{-2}$. To assess the temporal dependence, all data of the photoinduced intensity changes are fitted to the following materials response function convoluted with an instrumental response time of 2.0 ps:

$$\ln\frac{I(t)}{I_0} = A \cdot \left[1 - \exp\left(-\frac{t}{\tau_r^{\perp}}\right)\right] \cdot \exp\left(-\frac{t}{\tau_d}\right), \quad \text{(S1)}$$

where $A < 0$ is a constant related to the intensity decrease and $\tau_r^{\perp}$ and $\tau_d$ are the time constants for the rise and decay of the observed changes. Hence, the maximum intensity decrease at each laser fluence is obtained (shaded region in Fig. S4b, d, and f) for the fluence dependence shown in Fig. 3.

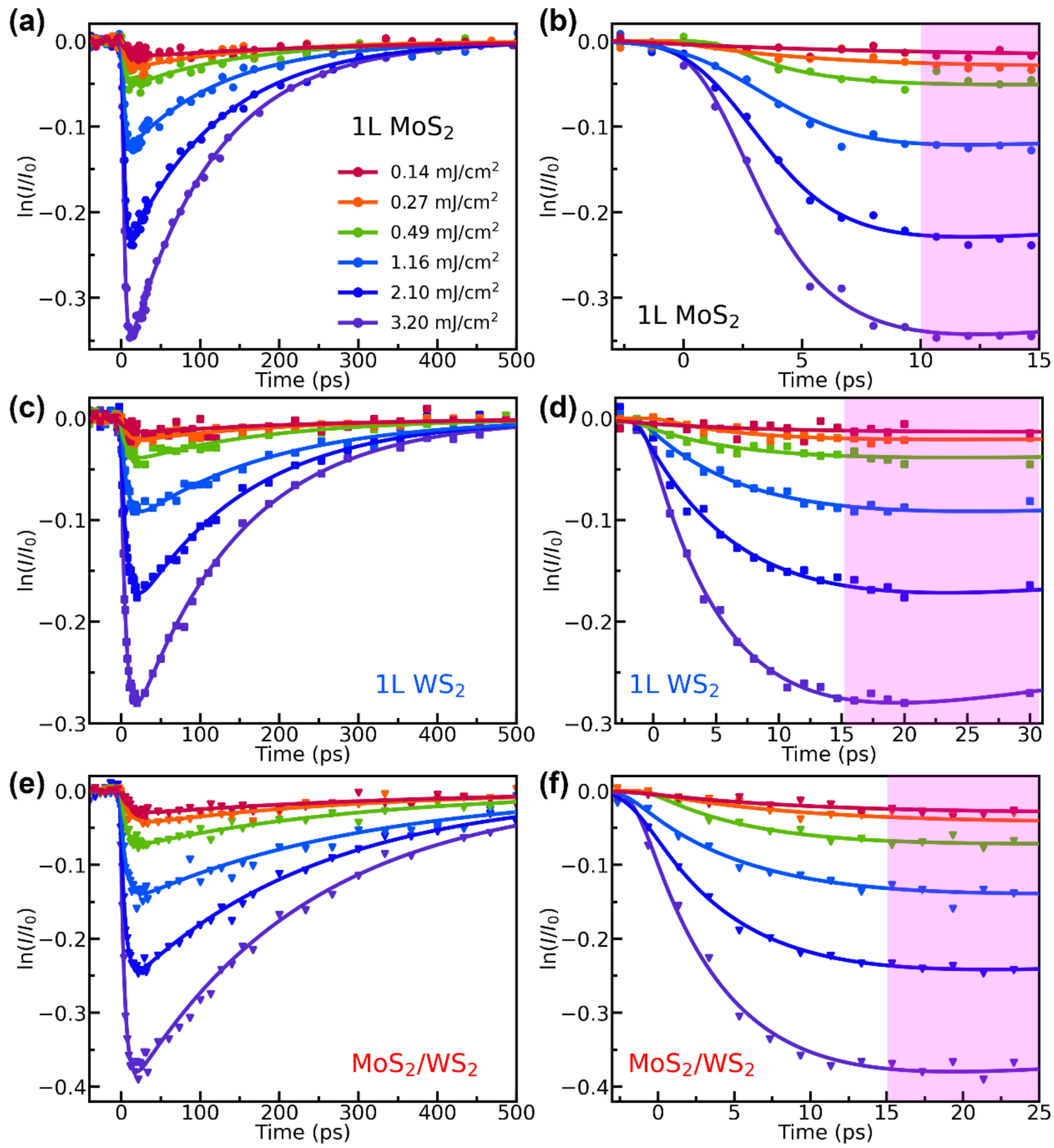

**Figure S4.** Structural dynamics of the TMDC monolayers and HS recorded at select laser fluences ranging from 0.14 to 3.20 mJ cm$^{-2}$. (a, c, e) Overall diffraction intensity changes for 1L $MoS_2$, 1L $WS_2$, and $MoS_2/WS_2$ HS, respectively. Solid lines are fits based on Eq. S1 convoluted with the instrumental response time. (b, d, f) The corresponding early-time dynamics. The shaded regions are used to obtain the maximum intensity decreases and their standard deviations.

## S4. Dependence of diffraction intensities on OP $A_1'$ or $A_2''$ motions

We calculate the structure factor to examine the effects of the OP $A_1'$ and $A_2''$ vibrational motions on the diffraction intensity of $MoS_2^{HS}$ in the HS. At equilibrium, the Mo atom is in the center of the unit cell and two S atoms are located at $\pm z_S$ =1.586 Å. Since the $A_1'$ mode is symmetric with opposite S atomic displacements $\pm\delta$, the corresponding structure factor is

$$F_1(s_\perp, \delta) = f_{\mathrm{Mo}}(s_\perp) + f_{\mathrm{S}}(s_\perp)\exp\big(-i2\pi s_\perp(z_{\mathrm{S}} + \delta)\big) + f_{\mathrm{S}}(s_\perp)\exp\big(-i2\pi s_\perp(-z_{\mathrm{S}} - \delta)\big)$$
$$= f_{\mathrm{Mo}}(s_\perp) + 2\, f_{\mathrm{S}}(s_\perp)\cos\big(2\pi s_\perp(z_{\mathrm{S}} + \delta)\big)$$

where $f_{\mathrm{Mo}}$ and $f_{\mathrm{S}}$ are the atomic scattering factors of Mo and S at the vertical momentum transfer of $s_\perp$ = 1.0 Å$^{-1}$. The $A_1'$-specific diffraction intensity changes, $I_1(\delta) = \Delta|F_1|^2/\left|F_{1,\mathrm{eq}}\right|^2$, are shown in Fig. S5a. As for the $A_2''$ mode, the displacement of the Mo atom $\beta\delta$ is in the opposite direction of those of the two S atoms $-\delta$ with $\beta$ = 0.668 to conserve the center of mass of the unit cell. Hence, the associated structure factor is

$$F_2(s_\perp, \delta) = f_{\mathrm{Mo}}(s_\perp)\exp(-i2\pi s_\perp \beta\delta) + f_{\mathrm{S}}(s_\perp)\exp\big(-i2\pi s_\perp(z_{\mathrm{S}} - \delta)\big)$$
$$+ f_{\mathrm{S}}(s_\perp)\exp\big(-i2\pi s_\perp(-z_{\mathrm{S}} - \delta)\big)$$
$$= f_{\mathrm{Mo}}(s_\perp)\exp(-i2\pi s_\perp \beta\delta) + 2\, f_{\mathrm{S}}(s_\perp)\cos\big(2\pi s_\perp(z_{\mathrm{S}} - \delta)\big)\exp(i2\pi s_\perp \delta)\,.$$

The $A_2''$-specific diffraction intensity changes, $I_2(\delta) = \Delta|F_2|^2/\left|F_{2,\mathrm{eq}}\right|^2$, are shown in Figure S5b.

Because our instrumental response time is larger than the sub-100-fs oscillation periods of the $A_1'$ and $A_2''$ modes, an average of the intensity variations covering both positive and negative $\delta$ values is expected for mode-specific displacements, which means no change for $A_1'$ motions or a slight intensity increase (or almost no change considering the instrumental sensitivity and noise level) for $A_2''$ motions. However, we do not find strong evidence such as a noticeable onset delay for intensity change in our UED results, which signifies the lack of long-lasting coherent $A_1'$- or $A_2''$-only motions following the ultrafast interlayer charge transfer. This is reasonable because of the anticipated fast phonon–phonon scattering in a TMDC monolayer.

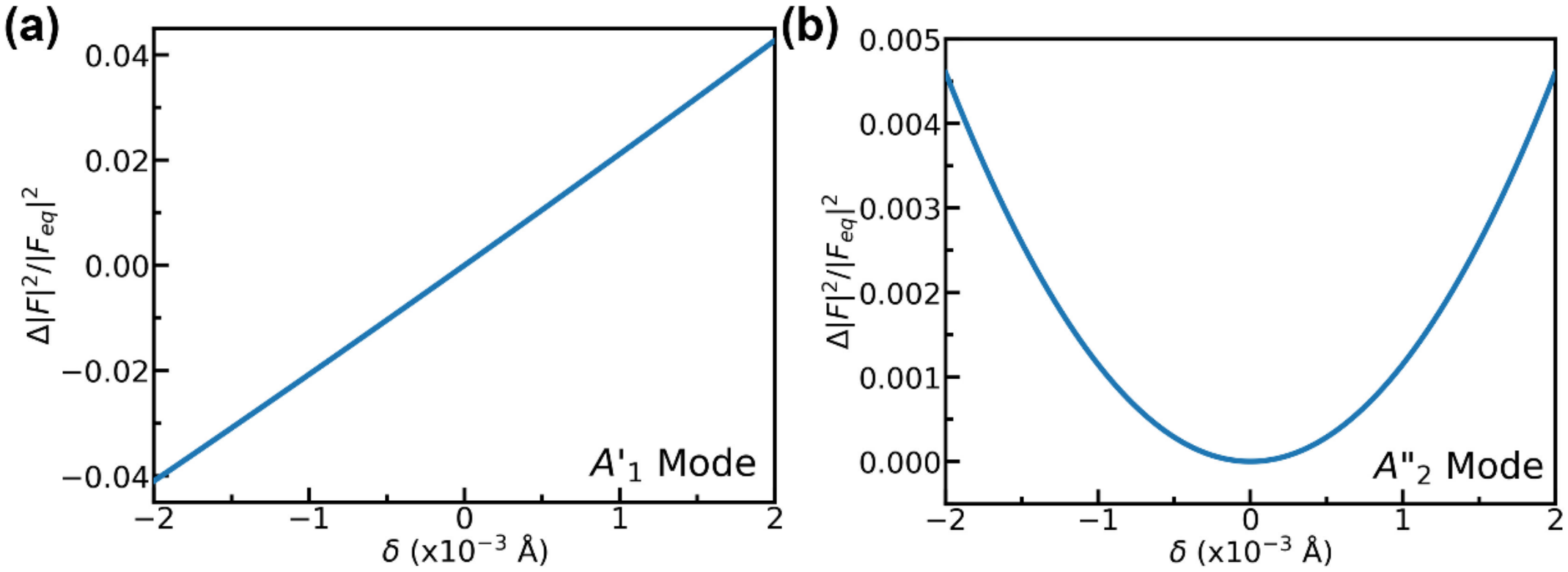

**Figure S5.** Diffraction intensity changes as a function of atomic displacement $\delta$ calculated based on coherent OP (a) $A_1'$ and (b) $A_2''$ motions of 1L $MoS_2$. The vertical momentum transfer is chosen to be $s_\perp$ = 1.0 Å$^{-1}$, which is close to where the experimental data are evaluated.

## S5. Analysis of optical absorption based on classical Fresnel equations

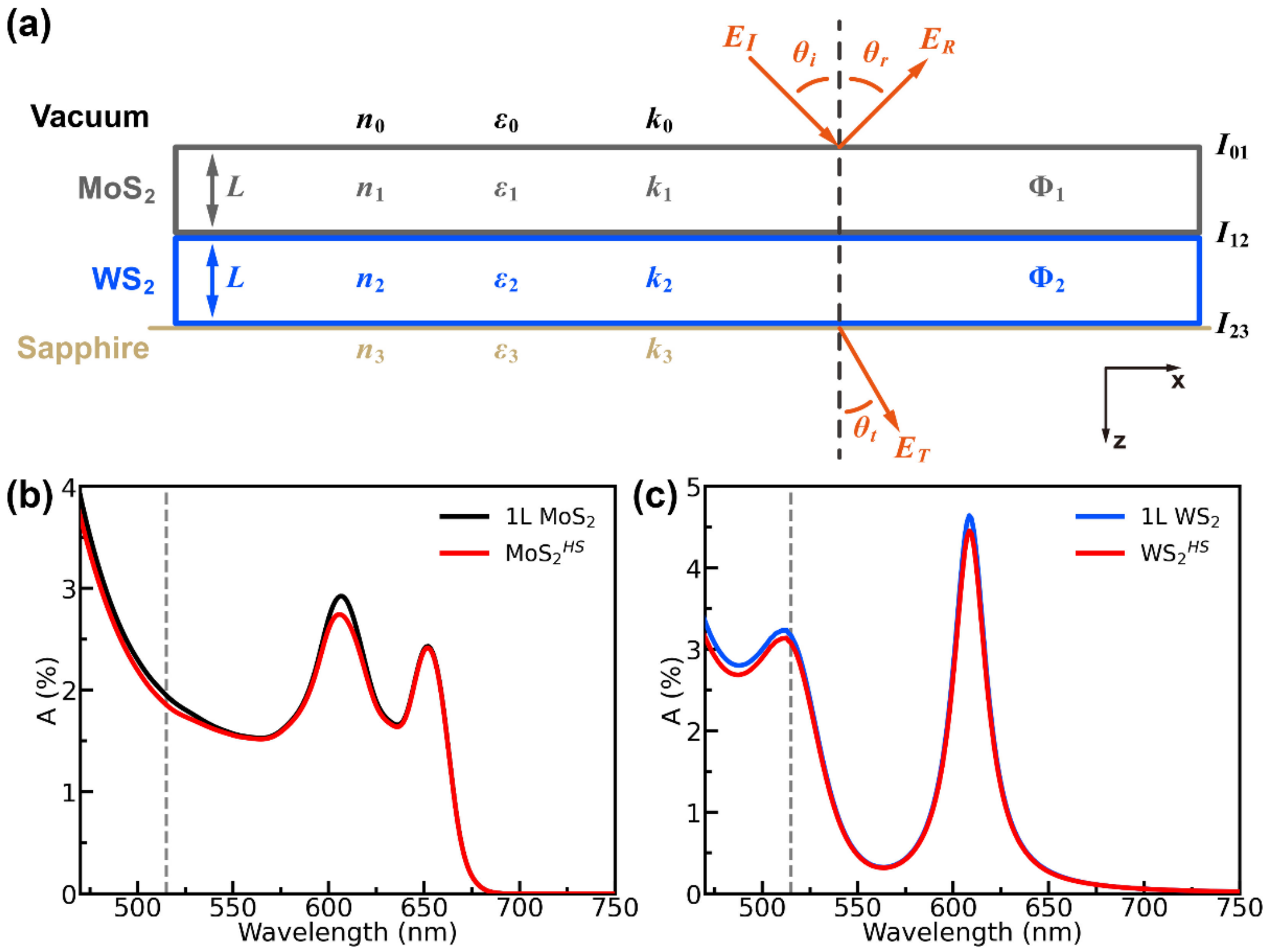

**Figure S6.** Schematic of the multilayer stack and the calculated absorption contributions for the $MoS_2/WS_2$ HS. (a) Consideration of a four-layer vacuum/$MoS_2$/$WS_2$/Sapphire system, with $E_I$, $E_R$, and $E_T$ denoting the incident, reflected, and transmitted beams. See text for details about the symbols involved. (b and c) Comparison of the absorption spectra of $MoS_2$ and $WS_2$ in separated (black or blue) and HS (red) conditions. The dashed line indicates 515 nm used for photoexcitation in this study.

We employ the transfer matrix method to calculate the classical optical response of a supported bilayer HS as a multilayer vacuum/$MoS_2$/$WS_2$/sapphire stack (see Fig. S6a for a schematic). Each layer $j$ (0 to 3) is characterized by its complex refractive index $\tilde{n}_j = n_j + ik_j$ and thickness $L_j$. When $p$-polarized light propagates across two adjacent layers at an incidence angle $\theta_i$ from $i$ to $j$, an interface matrix between the two layers is given by

$$I_{ij} = \frac{1}{t_{ij}} \begin{pmatrix} 1 & r_{ij} \\ r_{ij} & 1 \end{pmatrix},$$

where $r_{ij}$ and $t_{ij}$ are the Fresnel reflection and transmission coefficients given as follows:

$$r_{ij} = \frac{\varepsilon_j k_i - \varepsilon_i k_j}{\varepsilon_j k_i + \varepsilon_i k_j}, t_{ij} = \frac{2\varepsilon_i k_j}{\varepsilon_i k_j + \varepsilon_j k_i}$$

with $\varepsilon_{i,j} = \tilde{n}_{i,j}^2$being the complex dielectric constants and $k_{i,j} = 2\pi/\lambda\sqrt{\varepsilon_{i,j} - \sin^2\theta_i}$ being the OP wavevector components in the respective layers. Light propagation through layer $j$ is described by

$$\Phi_j = \begin{pmatrix} \exp(-ik_jL_j) & 0 \\ 0 & \exp(ik_jL_j) \end{pmatrix}.$$

Thus, the total transfer matrix $M$ for the four-layer system is obtained by the product

$$M = I_{01}\Phi_1 I_{12}\Phi_2 I_{23}\,.$$

The complex reflection and transmission coefficients for $p$-polarized light are obtained from the matrix elements of $M$ as

$$r_p = \frac{M_{11}}{M_{21}}\,, \qquad t_p = \frac{1}{M_{11}}$$

and therefore the reflectance and transmittance are given by

$$R_p = |r_p|^2\,, \qquad T_p = |t_p|^2 \frac{n_3 \cos\theta_t}{n_0 \cos\theta_i}\,,$$

where $\theta_t$ is the transmission angle in sapphire. The photoabsorption can then be calculated by $A_p = 1 - R_p - T_p$.

We use $\tilde{n}_0$ = 1.00 for vacuum, $\tilde{n}_3$ = 1.76 for sapphire,[S5] and $\tilde{n}_1$ and $\tilde{n}_2$ of the monolayers based on the reported experimental values.[S6,S7] For simplicity, a normal incidence ($\theta_i$ = 0°) is considered, and the thicknesses of both monolayers are taken to be $L$ = 0.70 nm.[S8] It is found that the overall absorption spectrum of the supported HS is similar to the combined spectral features of the two monolayers. However, the contribution of each layer to the absorption is actually slightly reduced compared to that in separation, as a result of multilayer interference and the redistribution of the optical field inside the stack (Fig. S6, b and c). The disagreement between the UED observations with clearly more photoabsorption and the classical Fresnel consideration further indicates the significance of interlayer electronic coupling in the optical behavior of HSs.

### S6. Thermal transport model for relaxation of laser-heated TMDC systems

The relaxation of laser-heated TMDCs is essentially one-dimensional (1D) because the electron-probed region is much smaller and well within the footprint of the laser-illuminated area. Therefore, we consider a 1D heat diffusion model (along $z$ into the sapphire bulk) together with thermal transport across the interfaces for a 2- (monolayer) or 3-component (HS) system. For the former, the model has been described in detail previously, where the recovery process is largely governed by the thermal conductance across a single interface from the laser-heated monolayer to the sapphire substrate.[S9] For bilayer HSs, an additional van der Waals heterojunction exists between the monolayers. Thus, thermal transport across two interfaces and in sapphire may be described by the following system of coupled equations:

$$C_{\mathrm{M}}\rho_{\mathrm{M}}L\frac{\partial T_{\mathrm{M}}}{\partial t} = -G_{\mathrm{HS}}(T_{\mathrm{M}}-T_{\mathrm{W}}),$$

$$C_{\mathrm{W}}\rho_{\mathrm{W}}L\frac{\partial T_{\mathrm{W}}}{\partial t} = G_{\mathrm{HS}}(T_{\mathrm{M}}-T_{\mathrm{W}}) - G_{\mathrm{W}}\big(T_{\mathrm{W}} - T_{\mathrm{S}}(z=0,t)\big),$$

$$-\kappa_{\mathrm{S}}\frac{\partial T_{\mathrm{S}}(z,t)}{\partial z}\bigg|_{z=0} = G_{W}\big(T_{\mathrm{W}} - T_{\mathrm{S}}(z=0,t)\big),$$

$$\frac{\partial^2 T_{\mathrm{S}}(z,t)}{\partial z^2} - \frac{C_{\mathrm{S}}}{\kappa_{\mathrm{S}}}\frac{\partial T_{\mathrm{S}}(z,t)}{\partial t} = 0,$$

where $C$, $\rho$, $L$, and $T$ are, respectively, the specific heat capacity, mass density, thickness, and time-dependent temperature; the subscripts M, W, and S stand for $MoS_2$, $WS_2$, and sapphire, respectively; $G_{\mathrm{HS}}$ and $G_{\mathrm{W}}$ are the thermal boundary conductances (TBCs) across the HS and the $WS_2$–sapphire interfaces, respectively; and $\kappa_S$ is the thermal conductivity of sapphire.

**Table S1:** Parameters used in the thermal transport model.

| Parameter (Unit) | $MoS_2$ | $WS_2$ | Sapphire |
|---|---|---|---|
| $C$ (J g$^{-1}$ K$^{-1}$) | 0.397[S9] | 0.256[S10] | 0.75 |
| $\rho$ (g cm$^{-3}$) | 5.06[S9] | 7.5[S10] | 3.98 |
| $\kappa$ (W m$^{-1}$ K$^{-1}$) | 2.0[S11] | 2.0[S12] | 32.5[S13] |
| $L$ (nm) | 0.7[S9] | 0.7 | 500 |

The initial condition of the sapphire substrate is $T_{\mathrm{S}}(z, t=0)$ = 295 K for all $z$. The initial value of $T_{\mathrm{M}}$ is derived from the UED results (i.e., the highest temperature at each fluence in Fig. 5c), which is ~12% higher than 1L $MoS_2$ (Fig. 5a). Hence, the initial $T_{\mathrm{W}}$ value is also set to ~12% higher than that of 1L $WS_2$ (Fig. 5b). We note that the extent of the enhancement in $T_{\mathrm{W}}$ does not strongly impact the calculated results given that $T_{\mathrm{M}}$ is notably higher than $T_{\mathrm{W}}$ and therefore the direction of thermal conductance is clearly established. In addition, negligible temperature increase is found for $z > 500$ nm of the sapphire bulk.

By varying the TBC values, numerical calculations are carried out to find the best matches with the experimental data over the large temporal window shown in Fig. 5. Specifically, the calculated results for the fluence of 3.2 mJ cm$^{-2}$ are shown in Fig. S7. The surface temperature increase of sapphire is found to be quite limited due to its high thermal conductivity (Fig. S7b).

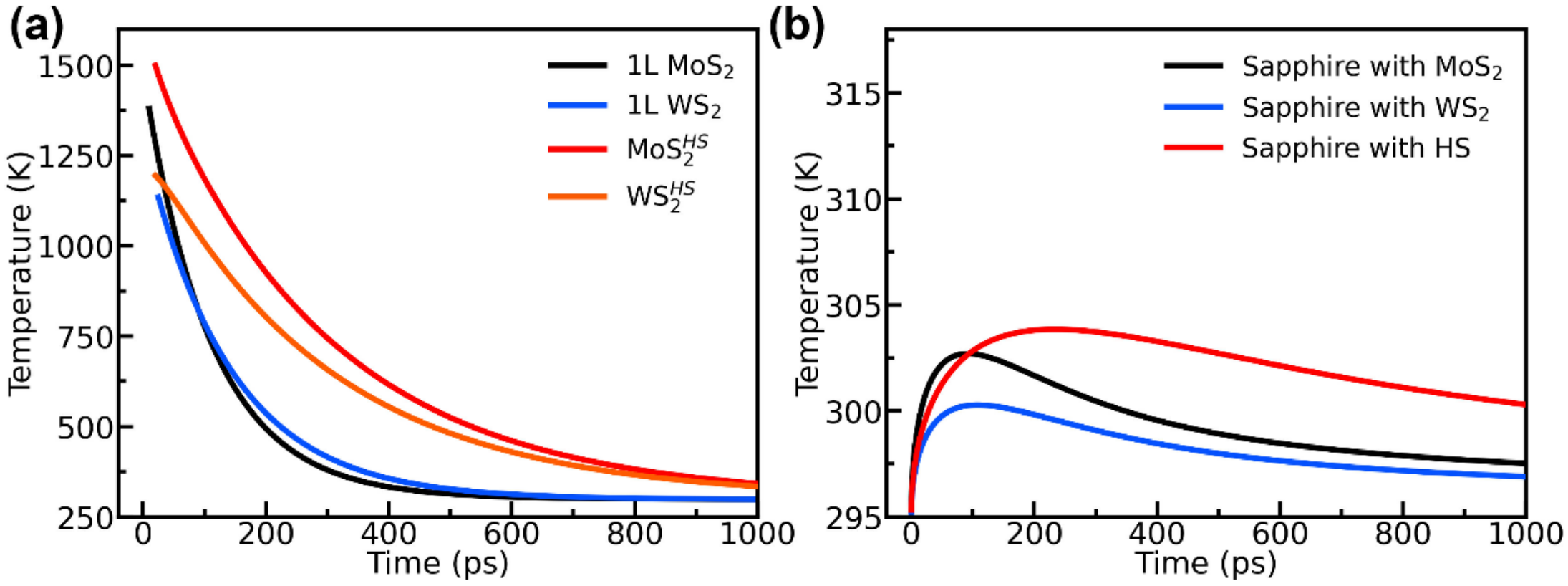

**Figure S7.** Thermal transport in supported 1L $MoS_2$, 1L $WS_2$, and the $MoS_2/WS_2$ HS laser-heated at 3.2 mJ cm$^{-2}$. (a) Calculated temperatures of the TMDC layers as a function of time. (b) Calculated temporal evolution of the sapphire surface temperature beneath 1L $MoS_2$ (black), 1L $WS_2$ (blue), and the HS (red).